# Deep Regression 2D-3D Ultrasound Registration for Liver Motion Correction in Focal Tumor Thermal Ablation

*Shuwei Xing[1,2], Derek W. Cool[4,5], David Tessier[1], Elvis C.S. Chen[1,2,3,4,5], Terry M. Peters[1,2,3,4,5] and Aaron Fenster[1,2,3,4,5]*

[1] *Robarts Research Institute, Western University, London, ON, N6A 5B7, Canada*
[2] *School of Biomedical Engineering, Western University, London, ON, N6A 5B7, Canada*
[3]*Department of Medical Biophysics, Western University, London, ON, N6A 5B7, Canada*
[4]*Department of Medical Imaging, Western University, London, ON, N6A 5B7, Canada*
[5]*Lawson Health Research Institute, London, ON, N6A 5B7, Canada*

**Abstract.** Liver tumor ablation procedures require accurate placement of the needle applicator at the tumor centroid. The lower-cost and real-time nature of ultrasound (US) has advantages over computed tomography (CT) for applicator guidance, however, in some patients, liver tumors may be occult on US and tumor mimics can make lesion identification challenging. Image registration techniques can aid in interpreting anatomical details and identifying tumors, but their clinical application has been hindered by the tradeoff between alignment accuracy and runtime performance, particularly when compensating for liver motion due to patient breathing or movement. Therefore, we propose a 2D-3D US registration approach to enable intra-procedural alignment that mitigates errors caused by liver motion. Specifically, our approach can correlate imbalanced 2D and 3D US image features and use continuous 6D rotation representations to enhance the model's training stability. The dataset was divided into 2388, 196 and 193 image pairs for training, validation and testing, respectively. Our approach achieved a mean Euclidean distance error of 2.28 $\pm$ 1.81mm and a mean geodesic angular error of 2.99 $\pm$ 1.95°, with a runtime of 0.22 seconds per 2D-3D US image pair. These results demonstrate that our approach can achieve accurate alignment and clinically acceptable runtime, indicating potential for clinical translation.

**Keywords:** Multi-modal image registration, tumor ablation, motion correction.

## 1 Introduction

Liver tumor ablation is an established therapeutic modality for the treatment of focal liver tumors [1], particularly in patients who are ineligible for surgical resection [2].

During ablation (radiofrequency or microwave) procedures, physicians usually require inserting a single needle-shaped applicator into the tumor centroid. Then a thermal ablation zone is generated surrounding the applicator tip for eradicating cancerous cells. While ultrasound (US) and computed tomography (CT) are both viable options for applicator guidance, US has advantages over CT due to its real-time imaging capabilities, lower cost and widespread availability [3]. However, this US-guided procedure relies heavily on the physician's experience to accurately place the applicator, as it lacks three-dimensional (3D) anatomical information for ensuring complete tumor coverage [4, 5]. Moreover, for some patients, the conspicuity of liver tumors in US images is low or almost non-existent [6]. In addition, some tumor mimics, such as regenerative nodules in cirrhotic liver and prior ablation sites may also confuse physicians [7], making tumor targeting task more challenging. Thus, US-CT/MRI registration and fusion techniques have been proposed and demonstrated to improve tumor visibility and reduce physician's variability concerning the interpretation of anatomical details [8]. However, the clinical application of these multi-modal fusion techniques has been hindered by the high clinical demands of the ablation procedure. Specifically, in addition to achieving clinically acceptable fusion accuracy, registration techniques must be capable of compensating for patient-related liver motion in real time, arising from patient breathing and occasionally irregular body movement [9].

To address these issues, image-based registration techniques have been extensively investigated, but existing straightforward solutions (directly from 2D US to CT or MRI) are challenging to be used clinically due to their high complexity. For example, to automate liver 2D US-CT/MRI alignment, the Linear Correlation of Linear Combination ($LC^2$) metric was introduced by Wein et al. [10]. However, its expensive computational cost still poses a challenge to allow real-time US guidance. Subsequently, Pardasani et al. [11] used a modified $LC^2$ metric to expedite the alignment in US-guided neurosurgical procedures, attaining a runtime performance accelerated by a PyCUDA-based framework resulting in approximately 5 fps. Recently, deep learning has been employed for this challenging registration problem, but current developments are still in their early stages [12].

To date, the standard solution for multi-modal registration is to introduce external tracking sensors. While Penney et al. [13] employed an optical tracker to obtain the spatial information of US images and proposed a US slice-to-MRI registration approach based on the probability map of corresponding structures. However, liver respiratory motion was not accounted for in this study. Wein et al. [9] attached a position-sensor to the patient's skin to detect anterior-posterior translation of the liver. Combined with a US slice-to-volume registration approach that used Local Normalized Cross Correlation (*LNCC*) as the similarity metric, they achieved approximately 5 fps performance but did not report the registration accuracy or robustness.

To simplify the multi-modal registration stage, 3D US imaging can be used to decompose the task into two sequential registration steps: "3D US-to-CT/MRI" [14, 15] and "dynamic 2D-to-3D US". This concept has been well demonstrated in prostate interventions [5, 16, 17]. For example, Xu et al. [16] demonstrated the feasibility of fusing transrectal US images with pre-procedural MRI images for prostate biopsy and used 3D US images as the "registration agent" to register with 2D transrectal US and MRI images. Additionally, Guo et al. [17] developed a deep learning-based US frame-to-3D image registration approach, which achieved real-time performance and did not require external tracking sensors. However, the feasibility of this concept for use in liver interventions is still under investigation due to the relatively large liver motion arising from patient breathing and irregular body movement.

Testing of our previously developed electro-mechatronic 3D US liver ablation guidance system in a 14-patient trial demonstrated that 3D US images could improve clinical outcomes [18-20]. Subsequently, we developed the first registration step, "3D US-to-CT/MRI", to facilitate the procedure [6]. Therefore, this work focuses on the second registration step, "dynamic 2D-to-3D US", to demonstrate the clinical effectiveness of mitigating the effect of liver motion, thereby improving tumor visibility during procedures. Specifically, we aimed to address the tradeoff between alignment accuracy and runtime for the "dynamic 2D-to-3D US" registration task. Thus, we proposed a deep learning-based 2D-3D US registration approach to allow accurate registration of a pre-procedural CT/MRI image with clearly visible tumors to the intra-procedural 2D US video stream and eliminate errors caused by liver motion. Our contributions include:

- A temporal 2D-3D US registration workflow that facilitates intra-procedure registration that is compatible with other registration algorithms.
- A deep regression dynamic 2D-3D US registration algorithm (DeepRS2V), using a dot-product operation-based module to combine imbalanced features between 2D and 3D US images, and avoid the discontinuity problem of rotation representations during training, via a 6D representation for transformation prediction.
- A clinically acceptable solution that achieves accurate alignment in close to real-time.

## 2 Method

### 2.1 Problem definition

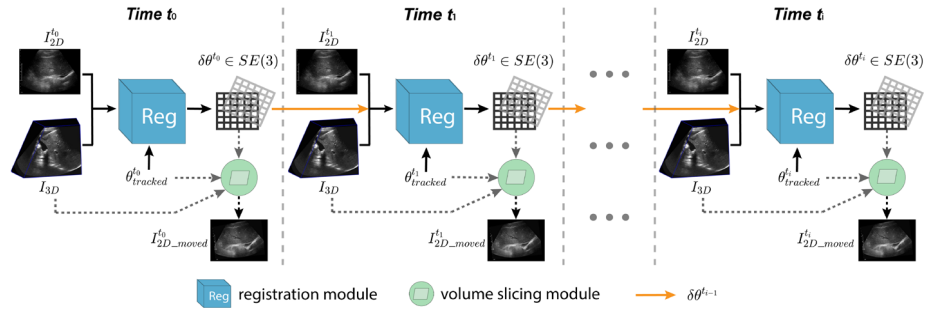

**Figure 1.** Workflow of temporal 2D-3D US registration. $t_0, t_1$ and $t_i$ represent different time points. $\theta_{tracked}$ and $\delta\theta$ are the transformation representations (translations + rotations).

A temporal 2D-3D US image registration workflow is proposed to facilitate the liver ablation guidance procedure. As shown in Fig. 1, the inputs at each time point $t_i$ include a reference 2D US image, $I_{2D}^{t_i}$ and a 3D US image, $I_{3D}$. Both were acquired using our developed electro-mechatronic 3D US liver system (see Fig. 2) [18]. The user can freely move the attached 2D US transducer at the end of the counterbalanced arm for acquiring real-time 2D US images [20]. 3D US images were reconstructed from a sequence of 2D US images acquired along a predefined trajectory using a motor-driven scanner, which is supported by our counterbalanced arm [21].

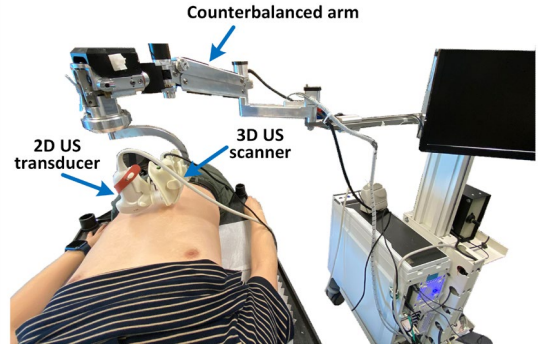

**Figure 2.** 3D US liver ablation guidance system. The 3D US scanner can automatically drive a conventional 2D US transducer along a predefined trajectory.

Since all joints in the arm of the tracking system have angular encoders, the relative transformation, $\theta_{tracked}^{t_i}$, from $I_{3D}$ to $I_{2D}^{t_i}$ can be easily obtained. Note that this transformation, $\theta_{tracked}^{t_i}$, cannot represent the internal liver motion. Thus, the objective of our registration module is to calculate the transformation, $\delta\theta^{t_i}$, for correcting the pose of the 3D US image to align with the 2D US image, which is affected by liver motion. Importantly, to account for the continuity of liver motion, the correction transformation, $\delta\theta^{t_{i-1}}$, obtained from the previous time point (arrows in orange in Fig. 1) is also used as the input to stabilize the registration process. Therefore, the 2D-3D US registration problem can be formulated as:

$$\delta\theta^{t_i} = \underset{\delta\theta}{\mathrm{argmin}}\, \mathcal{L}\{\mathcal{R}[I_{3D}, T(\theta_{tracked}^{t_i}, \delta\theta^{t_{i-1}})], I_{2D}^{t_i}\} \quad (1)$$

$$T(\theta_{tracked}^{t_i}, \delta\theta^{t_{i-1}}) = T(\theta_{tracked}^{t_i}) \cdot T(\delta\theta^{t_{i-1}}) \quad (2)$$

where $\mathcal{R}[I_{3D}, T]$ is the predicted 2D US image obtained by resampling $I_{3D}$ based on the transformation, $T(\theta_{tracked}^{t_i}, \delta\theta^{t_i})$, which corresponds to the "*volume slicing module*" (shown as the green circle in Fig. 1), and $\mathcal{L}\{*,*\}$ is the objective function (or similarity metric) that compares the predicted US image with the ground truth (i.e., the acquired real-time 2D US image). To use the best-performing "*registration module*", we developed a regression slice-to-volume registration algorithm-DeepRS2V-to address the clinical tradeoff between alignment accuracy and runtime.

## 2.2 DeepRS2V

For the 2D-3D US alignment, our hypothesis is that the correlation between encoded 2D US features $f_{2D}^{t_i}$ and 3D US features $f_{3D}$ can be learned and regressed to the transformation representations, $\delta\theta^{t_i}$, as shown in Fig. 3. The details of our DeepRS2V are as follows:

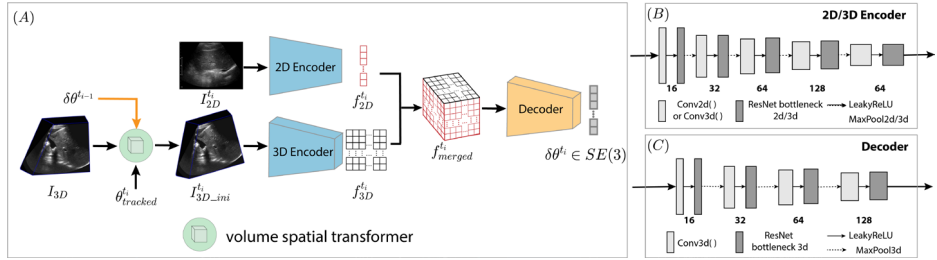

**Figure 3.** Architecture of DeepRS2V, demonstrating one instance of the "*registration module*" shown in Fig. 1.

**Volume Spatial Transformer** (or "*volume slicing module*" in Fig. 1). Unlike commonly used medical image processing toolkits (ITK, VTK, etc.), image metadata, such as the image origin, orientation, and spacing, cannot usually be retrieved to assist in the image spatial transformation of deep learning platforms. To apply the transformation to 3D US images and extract a 2D US slice, a differentiable module was designed based on the spatial transformer network [22]. First, given the transformation representations $\theta$ and $\delta\theta \in SE(3)$, a sampling grid $G_T$ is generated:

$$G_T = T(\theta, \delta\theta) \cdot G_0 \qquad (3)$$

where $T(*)$ is the 3 by 4 rigid transformation and $G_0$ is the standard 3D grid of the same size as $I_{3D}$. Then, image intensities are interpolated on grid $G_T$. In addition to obtaining the transformed volume as in the "*volume spatial transformer*", the "*volume slicing module*" can extract a 2D US slice (i.e., $I_{2D\_moved}$ in Fig. 1) from the transformed volume.

**2D-3D US Feature Fusion.** We used 2D and 3D encoders (see Fig. 3B) to extract low-level features $f_{2D}$ and $f_{3D}$, respectively. However, these encoded features $f_{2D}$ and $f_{3D}$ are highly imbalanced due to the difference in their image dimensions. To avoid the domination of features from $f_{3D}$ over $f_{2D}$, and to create feature correlation for the subsequent transformation regression task, a dot-product-based feature fusion operation was proposed, as shown in Eq. 4.

$$f_{merged}(i,j,k) = f_{3D} \cdot f_{2D} = \prod_{i,j}\prod_{k} f_{3D}(i,j) * f_{2D}(k) \qquad (4)$$

where $i, j$ and $k$ are the coordinates of the feature elements. The fused feature $f_{merged}$ becomes a "volume", which is decoded by a 3D decoder (see Fig. 3C) to predict the transformation representation.

**Transformation Representation.** In this work, a rigid transformation was used to demonstrate the feasibility of our proposed model and to investigate how well a simple rigid transformation corrects liver motion. Zhou et al. [23] defined the concept of continuous representation for rotation-based Euclidean topology, and they demonstrated that 6D and 5D representations (continuous) could outperform four or fewer dimensional representations (discontinuous), such as Euler angles, quaternions, and axis-angles, in pose estimation tasks. Inspired by that work, "6D rotation + 3D translation" representations were employed as our transformation representations. The rotation mapping from 3 × 3 matrices to a 6D representation was achieved simply by dropping the last column of the rotation matrix:

$$g_{GS}\left(\begin{bmatrix} | & | & | \\ a_1 & a_2 & a_3 \\ | & | & | \end{bmatrix}\right) = \begin{bmatrix} | & | \\ a_1 & a_2 \\ | & | \end{bmatrix} \qquad (5)$$

The mapping from a 6D representation back to a 3 × 3 rotation matrix is shown by Eq 6.

$$f_{GS}\left(\begin{bmatrix} | & | \\ a_1 & a_2 \\ | & | \end{bmatrix}\right) = \begin{bmatrix} | & | & | \\ b_1 & b_2 & b_3 \\ | & | & | \end{bmatrix}$$

$$b_i = \left[\begin{cases} N(a_1) & if\ i = 1 \\ N(a_2 - (b_1 \cdot a_2)b_1) & if\ i = 2 \\ b_1 \times b_2 & if\ i = 3 \end{cases}\right]^T \qquad (6)$$

where $N(\cdot)$ is a vector normalization function. All the operations are based on the Gram-Schmidt process.

**Loss function.** The $LNCC$ metric was used due to its robustness to US transducer orientation-induced variation in local brightness and contrast [9]. Since the size of predicted 2D US images varies during the registration process, which differs from the size of the 2D US reference image in our case (see Fig. 4), we modified the $LNCC$ to automatically calculate the metric on the intersection of the reference and predicted 2D US images. Thus, our customized $LNCC$ can handle any intersection shapes, including rectangular and polygonal. By doing so, our modified $LNCC$ can stabilize the model training even when small valid overlaps exist between 2D US reference and predicted US images.

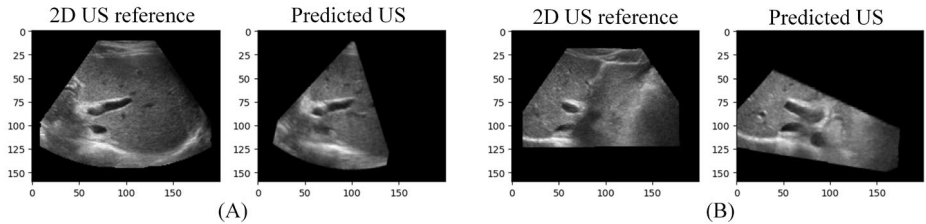

**Figure 4**. Comparison of 2D US reference image with the predicted US image.

Additionally, there were two supervised components needed to regularize the model training. One was the mean squared error (*MSE*) of the translational difference between the predicted translations, $\delta\theta_{trans}$, and the "ground truth (translations)", $\delta\theta_{trans}^{gt'}$. The other was the geodesic angular error between the predicted rotations, $\delta\theta_{rot}$, and the "ground truth (rotations)", $\delta\theta_{rot}^{gt'}$, as shown in Eq. 8. The geodesic error measures the minimal angular difference between two rotations [23]. Note that the "ground truth" is detailed in section 3.3 ("Registration evaluation"). The combined loss function is shown in Eq. 7:

$$\boldsymbol{\mathcal{L}} = \alpha \cdot LNCC + \beta \cdot \left\| \delta\theta_{trans} - \delta\theta_{trans}^{gt'} \right\|_2 + \gamma \cdot GeoErr(\delta\theta_{rot}, \delta\theta_{rot}^{gt'}) \quad (7)$$

$$GeoErr\left(\delta\theta_{rot}, \delta\theta_{rot}^{gt'}\right) = \cos^{-1}(\frac{M_{00}'' + M_{11}'' + M_{22}'' - 1}{2})$$
$$M'' = T(\delta\theta_{rot}) \cdot T\left(\delta\theta_{rot}^{gt'}\right)^{-1} \quad (8)$$

where factors $\alpha, \beta$ and $\gamma$ are used to balance the magnitude of the different metrics, with their sum equal to 1.0.

## 3  Experiments

### 3.1  Dataset

**Data acquisition.** Image data for this study were obtained from healthy volunteers at our institution under a Research Ethics Board-approved protocol. All subjects provided informed consent to the study. In our work, we used an iU22 US system with a C5-1 transducer (Philips, Eindhoven, Netherlands) to acquire images. We collected, on average, four different 3D US images for each participant to cover the entire liver, as shown in Fig. 5. 3D US image acquisition was performed during a 7-12s breath-hold, and the imaging depth was set at 14-18 cm. The sizes of 3D US images ranged from $708 \times 506 \times 253$ voxels to $708 \times 556 \times 278$ voxels, and voxel sizes were $0.17 \times 0.17 \times 0.33 \ mm^3$ to $0.32 \times 032 \times 0.65 \ mm^3$. For 2D US images, the physician freely scanned the liver under normal breathing conditions. The image sizes ranged from $752 \times 558$ pixels to $752 \times 564$ pixels and pixel sizes from $0.25 \times 0.25\ mm^2$ to $0.32 \times 0.32\ mm^2$. After excluding 2D-3D US image pairs with little or no overlap, our post-processed dataset included 1062 2D-3D US pairs (24 video clips) from 9 healthy volunteers.

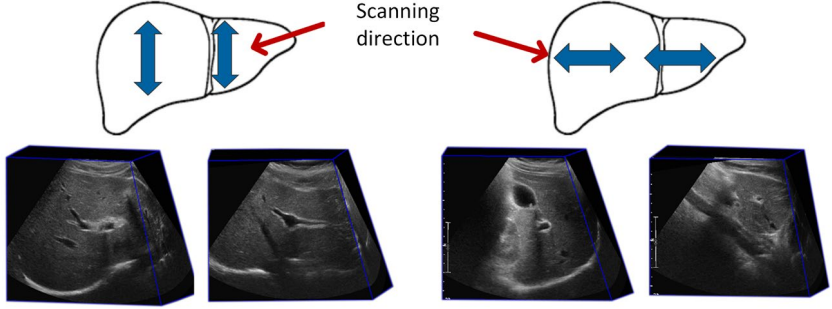

**Figure 5**. 3D US image acquisition showing how the US transducer was mechanically swept over the liver.

**Data augmentation.** We perturbed the initial transformation for each 2D-3D US image pair to augment the training dataset. For translation, $t$, uniform perturbations $u_{[-10,10]}, u_{[-10,10]}$ and $u_{[-5,5]}$ (in $mm$) were added to $t_x, t_y$, and $t_z$, respectively. For rotation, we added uniformly distributed angles $u_{[-5,5]}, u_{[-5,5]}$, and $u_{[-10,10]}$ (in degrees) to the Euler angles on the $x, y$, and $z$ axes. Note that the augmentation operations were only applied to the training dataset. After augmentation, the ratio of training, validation, and testing partitions was 2388: 196: 193 2D-3D US image pairs (or 14: 4: 5 video clips) respectively.

**Data preprocessing.** First, both 2D and 3D US images were isotopically resampled to $0.5\ mm$ spacing. Then, we center-cropped the 2D and 3D US images to sizes of $400 \times 320$ and $400 \times 320 \times 240$, respectively. Note that if the image was smaller than the cropping size, a zero-padding operation was applied to the edges. Lastly, image intensities were scaled to the range of 0 to 1.0. To ensure that only valid image area was considered, masks were applied to the 2D and 3D US images to remove regions outside of the valid image area.

### 3.2 Baseline approaches

To determine the best-performing "*registration module*", we compared DeepRS2V with an ITK-based approach and FVR-Net [24].

(1) *ITK-based approach*. The ITK-based approach was implemented using ITK registration packages. The global $NCC$ ($GNCC$) was used as the similarity metric with a regular gradient decent optimizer. The 2D and 3D US masks were applied to ignore the regions beyond the valid image area.

(2) The *FVR-Net*. The FVR-Net network was initially proposed by Guo et al. [24] for prostate biopsy procedures. A dual-branch balanced feature extraction module was used to combine the 2D and 3D US images. In FVR-Net, 3D US image features are extracted directly from the original 3D US image. Conversely, in our approach, these features are extracted from the initially transformed 3D US image (see $I_{3D\_ini}^{t_i}$ in Fig. 3). To adapt it to the liver case, our modified $LNCC$ was also used to re-train this FVR-Net model.

(3) *"DeepRS2V + correction"*. Based on DeepRS2V, we proposed a variant called "DeepRS2V + correction". For the "correction" part, we used the $LNCC$ and stochastic gradient descent ($SGD$) optimizer to further improve the alignment accuracy of DeeRS2V.

### 3.3 Registration evaluation

It is challenging to determine the ground truth when evaluating the registration accuracy. To address this problem, we first used our modified $LNCC$ as the similarity metric to register 2D and 3D US images, optimized by the $SGD$ optimizer. Next, we perturbed the obtained registration transformation by adding Gaussian-distributed noise $N(0, 1)$ and $N(0, 1.5)$ to translations and rotations, respectively. We generated 100 perturbed candidates (registered 2D US images) for each image registration pair. Lastly, the most visually similar image was chosen by a sonographer as the ground truth, primarily aiming to mitigate the limitation of $LNCC$ being trapped at a local minimum.

Additionally, the target registration error (TRE) was used to evaluate the registration accuracy using liver vessel bifurcation points chosen as landmarks. In 3D US images, vessel bifurcation points were selected based on the segmented 3D vessel surface model, which is described by Xing et al. [4]. In the 2D US setting, sequential 2D US images were acquired by freely sweeping over the human body by the physician. Thus, vessel bifurcation points can be localized on some 2D US images by visually examining the cognitively reconstructed vessels from sequential US images. Note that to reduce the impact of scanning speed on slice thickness, the sonographer manipulated the US transducer at a low speed.

### 3.4 Implementation details

Our registration model was implemented based on Pytorch[1] and the MONAI[2] framework was used for data preprocessing. The model was trained using the Adam optimizer with a starting learning rate of $10^{-6}$, which decayed by a gamma factor of 0.8 every 80 epochs. The training batch size and kernel size of $LNCC$ were set to 2 and 51, respectively. To balance each metric of the loss function, the values of $\alpha:\beta:\gamma$ were chosen as 20:1:10. We trained for 1200 epochs until convergence on an NVIDIA Quadro RTX 6000 and an RTX 2080 Ti, respectively. The source code is available at https://github.com/Xingorno/DeepRegS2V.

## 4 Results

### 4.1 Registration accuracy evaluation

To evaluate the registration accuracy, we compared our proposed approaches with ITK-based and FVR-Net approaches. 193 2D-3D US image pairs from 5 testing cases were used. Table 1 shows the 3D Euclidean distance error (in mm), which was also decomposed into the X, Y and Z directions (i.e., $T_x$, $T_y$, and $T_z$), as shown in Fig. 6. Additionally, the geometric angular error (in degrees) was used to calculate the 3D rotational error, which was also decomposed and reported relative to the X, Y and Z axes. Meanwhile, Table 1 shows the number of cases with a 3D Euclidean distance error of less than 10 mm labeled as "successful pairs". Lastly, the multiple comparison Dunnett's test [25], was used to analyze the statistically significant differences between the control approach ("DeepRS2V + correction") and other approaches.

**Table 1.** Registration pose error based on 193 2D-3D US image pairs. "Successful pairs" are the cases with a total translation error of less than 10 mm. The "DeepRS2V + correction" approach were used as the control group to calculate the $p$-values in a multiple comparison test.

| Methods | Translational error | | | | | Rotational error | | | | | Success-ful pairs |
|---|---|---|---|---|---|---|---|---|---|---|---|
| | $T_x(mm)$ | $T_y(mm)$ | $T_z(mm)$ | Euclidean distance($mm$) | $P$-value | $R_x(°)$ | $R_y(°)$ | $R_z(°)$ | Geometric angular(°) | $P$-value | |
| ITK-based | 1.09 | 0.97 | 1.89 | $2.70 \pm 1.62$ | $> 0.05$ | 1.89 | 2.16 | 1.16 | $3.50 \pm 2.34$ | $> 0.05$ | 192/193 |
| FVR-Net | 2.85 | 2.22 | 2.92 | $5.23 \pm 2.10$ | $< 0.05$ | 2.57 | 3.01 | 2.63 | $5.34 \pm 2.34$ | $< 0.05$ | 180/193 |
| DeepRS2V | 2.31 | 2.29 | 3.02 | $5.19 \pm 2.17$ | $< 0.05$ | 2.07 | 2.46 | 1.65 | $4.17 \pm 2.18$ | $< 0.05$ | 183/193 |
| DeepRS2V +correction | 0.97 | 0.96 | 1.40 | $\mathbf{2.28 \pm 1.81}$ | - | 1.64 | 1.78 | 0.93 | $\mathbf{2.99 \pm 1.95}$ | - | 190/193 |

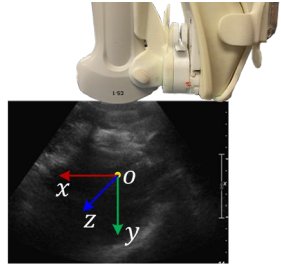

**Figure 6.** Definition of the coordinate system of 2D-3D US registration.

Table 1 shows that "DeepRS2V + correction" achieved a mean Euclidean distance error of $2.28 \pm 1.81$mm and a mean geodesic angular error of $2.99 \pm 1.95°$, and showed significant difference from the FVR-Net and DeepRS2V approaches. Additionally, the ITK-based approach also demonstrated low pose errors, but it required 10 s to 20s to register per 2D-3D image pair (see Table 3 below). DeepRS2V had the similar translational error to the FVR-Net, but a lower rotational error ($4.17 \pm 2.18°$ vs $5.34 \pm 2.34°$). In testing with 193 image pairs from 5 video clips, all approaches showed robustness with a success rate of over 95%, with a mean translation error of less than 10 mm, which were deemed to be successful.

For liver tumor ablation procedure, physicians typically use a 5 to 10 mm ablation margin beyond the detected boundary of the tumor to avoid residual tumors. Figure 7 shows the empirical cumulative distribution function of the translational error, demonstrating that more than 80% of image pairs have a mean Euclidean distance error of less than 5 mm, and all test cases (except for case 3) achieved a mean Euclidean distance error of less than 10 mm.

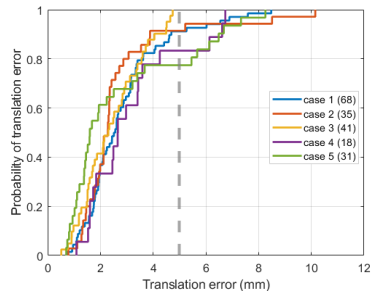

**Figure 7.** Cumulative distribution function of registration error on the testing cases. The number of image pairs is shown in legend brackets.

We also evaluated the TRE by calculating the distance between vessel bifurcation points in 2D and registered 3D US images. Except for case 4 with a TRE of $3.41 \pm 4.76 mm$, Table 2 shows that the TREs have a mean value of less than 1.5 mm, evaluated on at least 5 landmarks. In addition, Figure 8 shows the qualitative registration results for case 4 (with the worst TRE) and case 5 (with the best TRE) across different frames.

**Table 2.** TREs of the 5 testing cases

| Case # | Number of landmarks | TRE (mm) |
|---|---|---|
| 1 | 6 | $1.37 \pm 0.80$ |
| 2 | 5 | $1.40 \pm 1.86$ |
| 3 | 9 | $1.36 \pm 1.41$ |
| 4 | 5 | $3.41 \pm 4.76$ |
| 5 | 6 | $0.89 \pm 0.76$ |

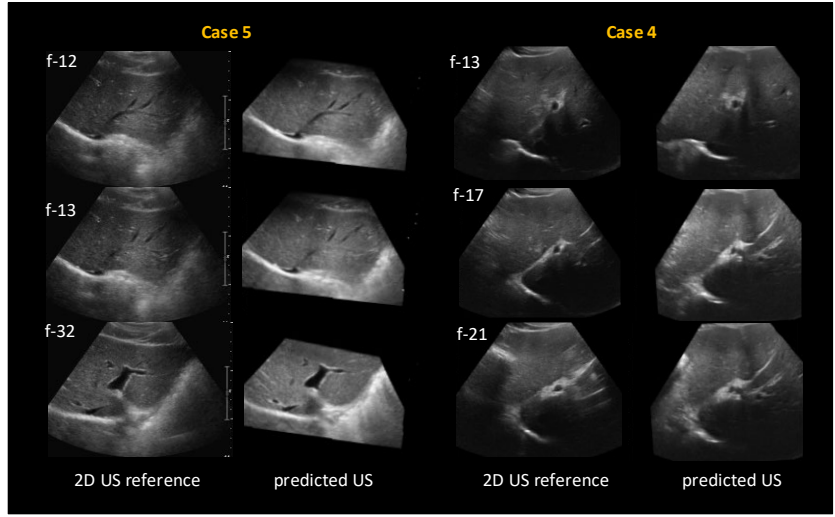

**Figure 8**. Registration results on cases 4 and 5 across different frames. Case 5 includes frames 12, 13 and 32, while case 4 includes frames 13, 17 and 21.

### 4.2 Registration runtime evaluation

For US-guided interventions, registration time is another critical aspect to demonstrate clinical applicability. Table 3 shows that DeepRS2V required an average of 0.10s and 0.05s to register a 2D-3D US image pair on an RTX 2080 Ti card and an RTX 6000 card, respectively. "DeepRS2V + correction" required 0.22s on an RTX 6000, which is slower than DeepRS2V, but still faster than other approaches. Testing on two different platforms, DeepRS2V and its variant achieved shorter registration time on an RTX 6000 card, indicating potential for further reducing the runtime on better hardware platforms.

**Table 3.** Registration runtime for each 2D-3D US image pair for different approaches on different platforms.

| Methods | Runtime (s) | | |
|---|---|---|---|
| | CPU(i7-9700k) | RTX 2080Ti | RTX 6000 |
| ITK-based | 10-20 | - | - |
| FVR-Net | - | 0.28 | 0.23 |
| DeepRS2V | - | 0.10 | 0.05 |
| DeepRS2V+correction | - | 0.37 | 0.22 |

### 4.3 Clinical integration

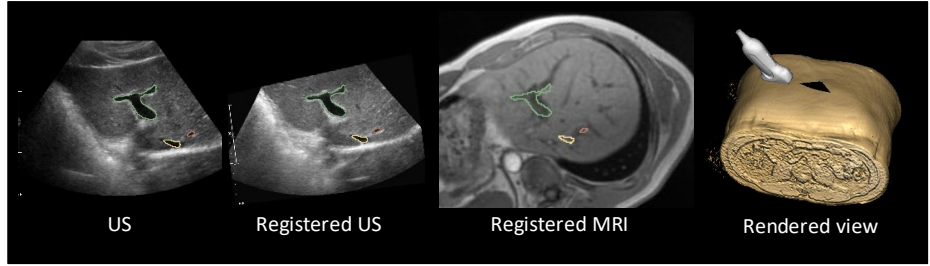

**Figure 9.** Registration result overview of 3 test cases. Segmented vessels in the 3D US image are overlaid on the 2D US image and registered with MRI images, respectively.

Figure 9 shows our approach in a clinical setting. Combined with a 3DUS-CT/MRI registration method developed by [6], the registered US and MRI images can be displayed simultaneously in real time during the procedure. The overlaid vessels on US and registered MRI images show the qualitative alignment performance, while the rendered 3D view provides the relative spatial relationship between the US transducer and the patient.

## 5 Discussion

The "DeepRS2V + correction" approach not only achieved the best performance in translation and rotation compared to other approaches, but also had a clinically acceptable runtime to facilitate US-guided procedures. For liver tumor ablation, an error of 5 mm in targeting the tumor centroid is deemed to be clinically acceptable, when considering a 5 to 10 mm typically safety margin [26]. Given that our resulting transformation is relative to the center of the 3D US images, a translation error of $2.28\ mm$ is adequate for tumor targeting. Additionally, the visualization of overlaid vessels and rendered views can further provide the physician with confidence in initiating treatment. A runtime of 0.22 s was achieved for each 2D-3D US image pair by "DeepRS2V + correction". Compared to other existing approaches, this provides smooth alignment in close to real-time, which is clinically acceptable.

In this work, a rigid transformation model was used to assess its effectiveness in correcting liver motion. As known, liver deformation can also occur due to patient breathing and movement. After testing 5 cases using rigid correction alone, the mean errors ($2.28 \pm 1.81 mm$ in translation and $2.99 \pm 1.95°$ in rotation) not only demonstrated the clinical feasibility of our approach, but also suggested that soft tissue deformation may not significantly impact this procedure. Previously, we discussed the impact of a rigid transformation model on the 3D US-CT/MRI registration task, which achieved a TRE of approximately 5 mm [6]. In contrast to the 2D-3D US registration performance, preliminary results indicate that developing deformation registration model for the 3D US-CT/MRI registration task may be more impactful than for 2D-3D US registration.

Our mechatronic arm tracking system not only provides the initial pose for stabilizing the registration, but also allows visualization of the relative spatial relationship between the patient and a US transducer. Since the acquired 3D US images have a limited field of view, it is possible during the procedure that some 2D US images may be beyond the 3D US imaging field. In such case, the 2D-3D US registration task may fail, but the relative relationship between the patient and a 2D US transducer can still function effectively, as shown in Fig. 9. This capability gives the physician confidence to proceed with the procedure.

## 6       Conclusion

We proposed a deep regression 2D-3D US registration algorithm embedded in a sequential registration workflow to correct liver motion in US-guided tumor ablation. The results demonstrated that our approach could readily address the tradeoff between registration accuracy and runtime, showing potential for clinical translation.